\documentstyle[preprint,aps,eqsecnum]{revtex}
\begin{document}
\draft
\tightenlines
\newcommand{\comm}[1]{\underline{\tt #1}}

\newcommand{\modegy}{\,({\rm mod}\,1)}
\newcommand{\atn}{\tan^{-1}}
\newcommand{\ch}{{\rm ch}}
\newcommand{\sh}{{\rm sh}}
\newcommand{\th}{{\rm th}}
\newcommand{\bml}{\begin{mathletters}}
\newcommand{\eml}{\end{mathletters}}
\newcommand{\be}{\begin{equation}}
\newcommand{\ee}{\end{equation}}
\newcommand{\ba}{\begin{array}}
\newcommand{\ea}{\end{array}}
\newcommand{\bea}{\begin{eqnarray}}
\newcommand{\eea}{\end{eqnarray}}
\newcommand{\nn}{\nonumber}
\newcommand{\smi}{\!-\!}
\newcommand{\spl}{\!+\!}
\newcommand{\pr}{\prime}
\newcommand{\dpr}{\prime\prime}
\newcommand{\cS}{{\Delta\cal S}}
\newcommand{\OL}{\overline}
\newcommand{\cK}{{\cal K}}
\newcommand{\cL}{{\cal L}}

\def\eqalign#1{
\null \,\vcenter {\openup \jot \ialign {\strut \hfil $\displaystyle {
##}$&$\displaystyle {{}##}$\hfil \crcr #1\crcr }}\,}

\pagestyle{myheadings}
\title{
{\flushright{\small cond-mat/0402129\\} \vspace{.2in}}
$O(1)$ contribution of saddle point fluctuations to the\\
free energy of Bethe Ansatz systems}
\author{\large F.~Woynarovich\footnote{e-mail: fw@szfki.hu}}
\address{Institute for Solid State Physics and Optics\\
Hungarian Academy of Sciences\\
1525 Budapest 114, Pf 49.}
\maketitle
\begin{abstract}
We develop a method to calculate the contribution of the
saddle-point fluctuations to the partition function of systems
soluble by the Bethe Ansatz. Using this method we give the $O(1)$
corrections to the free energy of the 1D repulsive $\delta$ Bose 
gas both for periodic boundary conditions and for the open end case. 
We also generalize our method to more complicated systems and discuss 
the case of XXZ Heisenberg chain in more details.
\end{abstract}


\setlength{\parskip}{2ex}
\setlength{\parindent}{0em}
\setlength{\baselineskip}{3ex}

\section{Introduction}
\label{sec:Int}
\noindent
Thermodynamic potentials such as the free energy are usually calculated in the
infinite size limit, and in this limit only the leading (macroscopic) contribution is taken, as in studying bulk or macroscopic properties this gives a sufficient accuracy.
Recent developments both in field theory and solid state physics have posed 
series of problems, where surface or impurity contributions are important 
\cite{AffLud,Kondo}. As this contributions are less than the macroscopic ones, 
the otherwise formidable problem of calculating the next to leading order terms is
set in focus.

For the one dimensional (1D) systems soluble by Bethe Ansatz (BA) the free energy 
is calculated following the method developed by C.N.~Yang and C.P.~Yang \cite{CNYCPY} 
for the $\delta$ Bose gas. 
The basic idea of this method is that through the density of the  
momenta (rapidities) an entropy can be defined enabling one to write up 
the free energy functional of the finite temperature system. 
Minimization of this functional with respect to the momentum density yields the most probable
distribution of the momenta and the value of the (macroscopic) free energy. In terms 
of the partition function, the minimizing of the free energy functional 
corresponds to finding the state entering with highest weight into the partition function
(saddle point). The contribution
of the states around this one (saddle-point fluctuations) is an $O(1)$ factor to
the partition function, i.e.~an $O(1)$ correction in the free energy (if not zero).
Our aim is to develop a convergent functional integral method to 
calculate this contribution, what is actually a finite size correction to the free energy.
  
First, to avoid special difficulties due to the BA, we write up, evaluate and discuss the convergence problems associated to the 
functional integral for the case of the \emph{free Fermi gas}. Using the repulsive $\delta$ Bose gas -- the simplest BA system -- as an
example, we generalize the procedure to the BA systems. We give also a general
expression for the contribution of the saddle point fluctuations in more complicated systems, and discuss the case of XXZ Heisemberg chain in more detail.

The quantity we calculate for the free Fermi and $\delta$ Bose gas is the grand canonical partition function and the thermodynamical potential defined by its logarithm (grand canonical potential). In a strict sense this latter is different from the free energy, nevertheless it is not unusual to name it so. Throughout the paper
also we use this certainly less precise, but hopefully not confusing name.

We consider systems at periodic boundary conditions (PBC). As in this case there
is no boundary, the na{\"\i}ve conception, which identifies the $O(1)$ corrections
as boundary contributions \cite{AffLud}
would lead to the conclusion, that for PBC there are no $O(1)$ corrections. 
Surprisingly we find this is true only for the case of free Fermi gas, but does not hold for the interacting systems. In the case of the repulsive $\delta$ Bose gas we calculate the contribution of the saddle-point fluctuations also for open end systems. We find that also in this
case the corrections are given by bulk properties (the energy of the particles 
as a function of the rapidities and by the kernel, i.e.~the derivative of the scattering phase shifts), and the nature of the boundary is reflected in the structure of the contributions.

We have to note, that our method works in the thermodynamic (infinite size) limit only, and as a consequence the
correction we find gives the so-called residual entropy in the zero temperature limit
\cite{AffLud}.
For finite systems the entropy $S$ tends to the logarithm of the ground-state degeneracy $\ln D$
as the temperature $T$ tends to zero. The residual entropy we find behaves differently:
its $T\to0$ limit is not connected to the ground-state degeneracy $D$ rather it depends on other 
properties like particle density or magnetization. This is a manifestation of the fact, that the ${\rm size}\to\infty$ and $T\to0$ limits do not commute.

The usual structure of the Bethe Ansatz equations makes it possible to give a general expression
for the contribution of the saddle point fluctuations in a large class of Bethe Ansatz systems.
Using the XXZ chain as an example we discuss some points one has to pay attention to when 
applying our formula. This considerations show that in addition to 
the saddle point fluctuations other effects can give $O(1)$ corrections too, more over in the isotropic (SU(2) symmetric) case the degeneracy of the states belonging to the same spin-multiplet
can give an even larger contribution. 

The paper is organized as follows. In Sec.~\ref{sec:fFg} we develop the method
and discuss the convergence problems. We use the free Fermi gas to have a possibility 
to check the result. The method is generalized  for the $\delta$ Bose gas in Sec.~\ref{sec:dBg}.
Here we treat both the PBC case and the case of open ends with surface potentials. 
Sec.~\ref{sec:Generalization} is devoted to the generalization of the method for more complicated 
systems. Our results are summarized in Sec.~\ref{sec:summary}. In the bulk of the paper we concentrate on the main line of our method, and the technicalities are collected in appendices.   

\section{The free Fermi gas}
\label{sec:fFg}
\noindent
The state of a free Fermion system is characterized by a set of wavenumbers
\be\label{wnfe}
k_i={2\pi\over L}J_i\,.
\ee
where $L$ is the size and at PBC the $J$ quantum numbers are integers. 
The energy of a state is 
\be\label{energy}
E=\sum_ie(k_i) \quad\quad\mbox{where}\quad\quad e(k)=\left(k^2-\mu\right)
\ee
with $\mu$ being the chemical potential, by which the 
required particle number $N$ is fixed \cite{muinYY}, and the grand canonical
partition function is   
\be\label{pf1}
Z=\sum_{\{k_i\}}\exp\left\{-\beta\sum_ie(k_i)\right\}\,.
\ee
Here the summation over $\{k_i\}$ means summation over all possible $k_i$ sets, and
$\beta$ is the inverse temperature $1/T$. 

Now we follow the method developed by C.N.~Yang and C.P.~Yang \cite{CNYCPY}
for the BA systems.
We split up the $k$ axis into intervals $\Delta k$, and introduce the 
density of $k$s ($\rho(k)$) and holes ($\rho_h(k)$) so that the number of the wavenumbers and holes
in the $(k,k+\Delta k)$ interval is given by $L\Delta k\rho(k)$ and 
$L\Delta k\rho_h(k)$ respectively. Obviously 
\be\label{rorohfe}
\rho(k)+\rho_h(k)={1\over2\pi}\,.
\ee
The number of states characterized by the same $\rho(k)$ function is
\be
\Omega\left[\rho(k)\right]=\prod_k\omega(\rho(k))\,,\quad{\rm with}\quad
\omega(\rho(k))={L\Delta k(\rho(k)+\rho_h(k))\choose L\Delta k\rho(k)}\,,
\ee
where in $\prod_k$ $k$ labels the $\Delta k$ intervals.
In the following $\ln\omega(\rho(k))$ is calculated by Stirling's
formula 
\be\label{lnomegaro}
\ln\omega(\rho(k))=L\Delta ks(\rho(k))+\varsigma(\rho(k))
+O\left({1\over L\Delta k\rho(k)},{1\over L\Delta k\rho_h(k)}
\right)
\ee
where
\be
s(\rho(k))=\left\{(\rho(k)+\rho_h(k))\ln(\rho(k)+\rho_h(k))-
\rho(k)\ln\rho(k)-\rho_h(k)\ln\rho_h(k)\right\}\,,
\ee
and
\be
\varsigma(\rho(k))=-{1\over2}\ln(2\pi L\Delta k)+{1\over2}\ln{\rho(k)+\rho_h(k)\over
\rho(k)\cdot\rho_h(k)}\,.
\ee
By means of $\omega(\rho(k))$ a free energy functional is defined
\be\label{fef}
F\left[\rho(k)\right]=\sum_k\left(
Le(k)\rho(k)\Delta k-T\ln\omega(\rho(k))\right)\,,
\ee
by which 
\be\label{pf2}
Z=\sum_{\{\rho(k)\}}e^{-\beta F\left[\rho(k)\right]}\,.
\ee
(Here the summation extends over all possible $\rho(k)$ distributions.)
In the usual procedure $\ln\omega(\rho(k))$ is calculated up to leading order in $L$ only, i.e.
\be
\ln\omega(\rho(k))\cong L\Delta ks(\rho(k))
\ee
is taken, 
and the free energy functional 
\be\label{fefL}
F_L\left[\rho(k)\right]=L\sum_k\left(
e(k)\rho(k)-Ts(\rho(k))\right)\Delta k
\ee
is minimized with respect to $\rho(k)$
leading to  
\be
{\rho_0(k)\over \rho_{h,0}(k)}=e^{-\beta e(k)}
\ee
and
\be
F_{\rm min}=-{L\over\beta2\pi}\sum\ln\left(1+e^{-\beta e(k)}\right)\Delta k=
-{L\over\beta2\pi}\int_{-\infty}^{\infty}
\ln\left(1+e^{-\beta e(k)}\right)dk\,.
\ee

Now we can define the functional integral for $Z$. 
We evaluate the summation over the possible $\rho(k)$ distributions through integrals over
the $\rho(k)$s: as $L\Delta k\rho(k)$ (being the number of $k$s in the interval $\Delta k$)
is integer, the correct approximation for the summs is  
\be
\sum_{\{\rho(k)\}}\Longrightarrow\int\cdots\int\prod_k\left(L\Delta kd\rho(k)\right)\,.
\ee
Introducing
\be
r(k)=\rho(k)-\rho_0(k)
\ee
we can expand $F\left[\rho(k)\right]$ around $\rho_0(k)$.
In order to have the functional integral convergent, in $\omega(\rho(k))$ of 
(\ref{lnomegaro}) we have to take into
account the terms next to the leading ones too ($\varsigma(\rho)$).
This gives
\bea
&&Z=e^{-\beta F_{\rm min}}\times\\
&&\int\cdots\int\prod_k\left(L\Delta kdr(k)\right)
\exp\left\{-\sum_k\left(L\Delta k{1\over2}{\rho_{0}+\rho_{h,0}\over
\rho_{0}\rho_{h,0}}\left(r(k)\right)^2-
{1\over2}\ln{\rho_{0}+\rho_{h,0}\over
\rho_{0}\rho_{h,0}L\Delta k2\pi}\right)\right\}\,.\nn
\eea
Here in the exponent the first term comes from the expansion of 
$F_L\left[\rho_0+r\right]$, and the second term is $\varsigma(\rho_0)$.
Introducing new variables
\be\label{xik}
\xi(k)=\sqrt{L\Delta k
{1\over2}{\rho_{0}(k)+\rho_{h,0}(k)\over
\rho_{0}(k)\rho_{h,0}(k)}}r(k)
\ee
one arrives at
\be
Ze^{\beta F_{\rm min}}=
\prod_k\left({1\over\pi}\int e^{-(\xi(k))^2}d\xi(k)\right)=1\,,
\quad\quad\mbox{i.e.}\quad\quad
Z=e^{-\beta F_{\rm min}}\,.
\ee

We can build up confidence in this calculation, as it reproduces the exact result.
As, however, we want to generalize the method to cases where no exact results
exist, it is worth to examine it in detail: it involves several approximations and also an implicit cut-off procedure, and we should see the conditions under 
which these are valid. Actually we have to check two types of approximations: 
\begin{enumerate}
	\item \emph{The application of the free energy functional.} The question to be answered is whether   the definition of the free energy functional is accurate enough. The partition functions 
	(\ref{pf1}) and (\ref{pf2}) are certainly equivalent in leading order, but are 
	not exactly equal, as (\ref{pf2}) is obtained from (\ref{pf1}) by a kind of averaging: 
	in each $\Delta{k}$ interval we take
	\be\label{summationapproximation}
	\sum_{\{k_j\}}\exp\left\{-\beta\sum_{k_j}e(k_j)\right\}
	\longrightarrow   
	\exp\left\{-\beta e(\overline{k})L\rho(k)\Delta{k}+\ln\omega(\rho(k))\right\}\,,
	\ee
with $\sum_{\{k_j\}}$ meaning summation over all possible choices of $L\rho(k)\Delta{k}$ $k_j$s out of the $L(\rho(k)+\rho_h(k))\Delta{k}$ possibilities, and $\overline{k}$ being a $k$ value within the interval $\Delta{k}$. As we are interested in calculating corrections to the macroscopic free
energy we have to define this averaging more precisely. In Appendix \ref{sec:errorinfeef} we show,
that taking for $\overline{k}$ the mean value of the $k$s in $\Delta{k}$ we introduce an
$O\left((\Delta{k})^2\right)$ error to the free energy \emph{density}, that disappears when the 
$\sum(\ldots)\Delta{k}\longrightarrow\int(\ldots)dk$ limit is taken. In a strict sense the mean value of the $k$s in $\Delta{k}$ still depends on $\rho(k)$ and $\rho_h(k)$, but in the practice 
one may take the middle of $\Delta{k}$ as $\overline{k}$. Although this introduces an $O(1/L)$
uncertainty in the mean value of energy/particle in each $\Delta{k}$ interval, as however, these are random, do not sum up to an $O(1)$ contribution. 

  \item \emph{The approximations applied while evaluating} (\ref{pf2}): 
  \begin{itemize}
     \item \emph{Conditions for the application of Stirling's formula.} Each $\xi$ integral collects the main part of the contribution from a region 
$-X<\xi<X$ where $X$ is a number of the order of 5-6 (\cite{AbSteg}).
We may use Stirling's formula in the above form, if in the corresponding $\rho$
region both $L\Delta k\rho(k),L\Delta k\rho_h(k)\gg1$. 
These lead to the requirements
\be\label{estimation1}
L\Delta k\rho_0(k)\left(1-X\sqrt{{2\rho_{h,0}(k)\over\rho_{0}(k)+\rho_{h,0}(k)}
\,{1\over L\Delta k\rho_0(k)}}\right)\gg1\ee
and
\be\label{estimation2}
L\Delta k\rho_{h,0}(k)\left(1-X\sqrt{{2\rho_{0}(k)\over\rho_{0}(k)+\rho_{h,0}(k)}
\,{1\over L\Delta k\rho_{h,0}(k)}}\right)\gg1\,,
\ee
i.e.~both condition are satisfied if 
\be\label{conditions}
L\Delta k\rho_0(k)\gg1\quad {\rm and} \quad L\Delta k\rho_{h,0}(k)\gg1\,.
\ee
As for large enough $k$
$\rho_0(k)\longrightarrow0$, one has to introduce a cutoff (say $\Lambda$) in the 
$k$ space so that (\ref{conditions}) are met for all $\vert k\vert<\Lambda$, i.e.
\be\label{acondition}
L\Delta k\rho_0(\Lambda)\gg1\,. 
\ee
This imposes a relation on the $L$ and $\Lambda$. 
Similarly, if the temperature $T$ is small enough
$\rho_{h,0}(0)\longrightarrow0$ thus the method is applicable at large enough size
but not at exactly zero temperature: the larger the size of the system is, the nearer the $T=0$
can be approached. 

We have to note, however, that since the entropy part of the \emph{macroscopic} free energy functional is already obtained through Stirling's formula, the  
problem connected to the application of this approximation is in principle present in the calculation of the macroscopic part of the free energy too (Appendix \ref{sec:lambdacutoff}).   
      \item \emph{Limits of the $\xi$ integrals.} The estimations 
(\ref{estimation1}-\ref{estimation2}) show also, that in case of (\ref{conditions})
the limits of the $\xi$ integrals can be 
taken to $\pm\infty$. 
     \item \emph{The $\varsigma(\rho_0)$.} We have taken $\varsigma(\rho)$ at $\rho_0$
and we neglected $\varsigma^{\prime}(\rho_0)r+\varsigma^{\prime\prime}(\rho_0)r^2/2$.
The term linear in $r$ shifts the center of the $\xi$ integral, while the
quadratic term modifies the coefficient of $\xi^2$. It is not hard to see that 
these corrections in the exponent are 
$O\left({1/L\Delta k\rho_0(k)},{1/L\Delta k\rho_{h,0}(k)}\right)$
ones, what we may neglect if (\ref{conditions}) holds. 
  \end{itemize}
\end{enumerate}
Now we may conclude, that our functional integral method is established in a 
strict mathematical sense in the $L\to\infty$ limit only, and involves a cutoff procedure
in the momentum space. 
We discuss the relation of the cutoff to the size and some details of the cutoff procedure (paying special attention to its connection to the macroscopic free energy) in Appendix \ref{sec:lambdacutoff}.

This method is not applicable for $T=0$, but the $T\to0$ limit is meaningful.
(In the present case the $T\to0$ limit reproduces the $T=0$ result, but this is not so in the case of interacting systems, as we shall see later.)

\section{The repulsive $\delta$ Bose gas}
\label{sec:dBg}
\noindent
\begin{flushleft}
\emph{The case of periodic boundary conditions}
\end{flushleft}
The Hamiltonian defining the 1D repulsive $\delta$ Bose gas is 
\be\label{deltaHam}
{\hat H}=-\sum_i{\partial^2\over\partial x^2_i}+2c\sum_{i<j}\delta(x_i-x_j)\,,\quad
c>0\,.
\ee
Using the Bethe Ansatz Lieb and Liniger has shown, that the diagonalization of 
this Hamiltonian in a box with a length of period $L$ can be reduced to the 
solution of the system of algebraic equations
\be\label{LiLie}
Lk_i=2\pi J_i-\sum_j^N 2\atn\left({k_i-k_j\over c}\right)\,,\quad J_i={N-1\over2}\modegy\,,
\ee
with $N$ being the number of particles in the system \cite{LiLi}. The 
energy of the system is again given by (\ref{energy}), but now the wavenumbers are 
determined by the Eqs.(\ref{LiLie}) instead of Eq.(\ref{wnfe}). The finite temperature 
description \cite{CNYCPY} goes like we treated the free electron gas with the difference, that  
due to Eq.(\ref{LiLie}) Eq.(\ref{rorohfe}) is replaced by
\bea\label{rorohdBg}
\rho(k)+\rho_h(k)&=&\sigma+\sum_{k^{\prime}}
K(k,k^{\pr})\rho(k^{\prime})\Delta k^{\prime}\nonumber\\
&=&\sigma+\int_{-\infty}^{\infty}
K(k,k^{\pr})\rho(k^{\prime})dk^{\prime}
\eea
with
\be
\sigma={1\over2\pi}\ \ \ \ \mbox{and}\ \ \ \ K\left(k,k^{\prime}\right)=
{1\over2\pi}\,{2c\over c^2+\left(k-k^{\prime}\right)^2}\,.
\ee
One should note, that only those $\rho(k)$ distributions are meaningful (physical), for which
this equation yields $\rho_h(k)\geq0$ for all $k$. The free energy is again given by (\ref{fefL}), but now it is constrained by (\ref{rorohdBg}). Its minimization leads to
\be
{\rho_0(k)\over \rho_{h,0}(k)}=e^{-\beta\epsilon(k)}\,,
\ee
with the energy $\epsilon(k)$ determined by the equation
\bea\label{epsilon}
\epsilon(k)&=&e(k)-{T}\sum_{k^{\prime}}
K(k,k^{\prime})\ln\left(1+
e^{-\beta\epsilon(k^{\prime})}\right)\Delta k^{\prime}\nn\\
&&\\
&=&e(k)-{T}\int_{-\infty}^{\infty}
K(k,k^{\prime})\ln\left(1+
e^{-\beta\epsilon(k^{\prime})}\right)dk^{\prime}\,.\nn
\eea
Once $\epsilon(k)$ is found, $\rho_0(k)$ is given by the equation
\be\label{rho0}
\rho_0(k)={1\over1+e^{\beta\epsilon(k)}}\sigma+\int_{-\infty}^{\infty}
{1\over1+e^{\beta\epsilon(k)}}K(k,k^{\pr})\rho_0(k^{\prime})dk^{\prime}\,,
\ee
and the minimal free energy is 
\be
F_{\rm min}=-{L\over\beta2\pi}\sum_k\ln\left(1+e^{-\beta\epsilon(k)}\right)\Delta k=
-{L\over\beta2\pi}\int_{-\infty}^{\infty}
\ln\left(1+e^{-\beta\epsilon(k)}\right)dk\,.
\ee
To continue we expand the (\ref{fefL}) around $\rho_0(k)$ and $\rho_{h,0}(k)$:
\be
F_L\left[\rho(k)\right]\simeq F_{\rm min}-T\sum_k L\Delta k{1\over2}
\left({\left(r(k)+r_h(k)\right)^2\over\rho_0(k)+\rho_{h,0}(k)}-
{r_h^2(k)\over\rho_{h,0}(k)}-
{r^2(k)\over\rho_0(k)}\right)\,.
\ee
Here the quantities 
\be
r(k)=\rho(k)-\rho_0(k)\quad{\rm and}\quad r_h(k)=\rho_h(k)-\rho_{h,0}(k)
\ee
are constrained due to (\ref{rorohdBg}) by 
\be\label{rrhdBg}
r(k)+r_h(k)=\sum_{k^{\prime}}K\left(k,k^{\prime}\right)r(k^{\prime})
\Delta k^{\prime}\,.
\ee
This way the functional integral for the partition function reads
\bea\label{funcintdBg}
&&Z=e^{-\beta F_{\rm min}}
\int\cdots\int\prod_k\left(L\Delta kdr(k)\right)\times\\
&&\exp\left\{-\sum_k\left(L\Delta k{1\over2}
\left({\left(r(k)+r_h(k)\right)^2\over\rho_0(k)+\rho_{h,0}(k)}-
{r_h^2(k)\over\rho_{h,0}(k)}-
{r^2(k)\over\rho_0(k)}\right)-
{1\over2}\ln{\rho_{0}+\rho_{h,0}\over
\rho_{0}\rho_{h,0}L\Delta k2\pi}\right)\right\}\,,\nn
\eea
what is to be evaluated under the constrain (\ref{rrhdBg}).
After eliminating $r_h(k)$ we arrive at
\bea
&&\sum_k L\Delta k{1\over2}
\left({\left(r(k)+r_h(k)\right)^2\over\rho_0(k)+\rho_{h,0}(k)}-
{r_h^2(k)\over\rho_{h,0}(k)}-
{r^2(k)\over\rho_0(k)}\right)=\nn\\
&&\nn\\
&&-{1\over2}\sum_{k,k^{\pr},k^{\dpr}}
L\Delta kr(k)K(k^{\pr},k){\rho_0(k^{\pr})\Delta k^{\pr}\over
\rho_{h,0}(k^{\pr})\left(\rho_0(k^{\pr})+\rho_{h,0}(k^{\pr})\right)}
K(k^{\pr},k^{\dpr})r(k^{\dpr})\Delta k^{\dpr}\nn\\
&&\\
&&+\sum_{k,k^{\pr}}
L\Delta kr(k){K(k,k^{\pr})\over\rho_{h,0}(k)}r(k^{\pr})\Delta k^{\pr}\nn\\
&&\nn\\&&\nn\\
&&-{1\over2}\sum_{k}L\Delta kr^2(k){\rho_0(k)+\rho_{h,0}(k)\over
\rho_0(k)\rho_{h,0}(k)}\,.\nn
\eea
Now, just as in the case of free Fermi gas we introduce
$\xi(k)$ according to Eq.(\ref{xik}). This leads to
\be
-\sum_{k,k^{\pr},k^{\dpr}}\xi(k)\left(\delta_{k^{\pr},k}-M_{k^{\pr},k}\right)
\left(\delta_{k^{\pr},k^{\dpr}}-M_{k^{\pr},k^{\dpr}}\right)\xi(k^{\dpr})\,,
\ee
where $\xi(k)$ is that of (\ref{xik}), $\delta_{k,k^{\pr}}$ is the Kronecker symbol, and
\be
M_{k,k^{\pr}}={1\over\rho_{h,0}(k)}\sqrt{{\rho_{0}(k)\rho_{h,0}(k)
\Delta k\over\rho_{0}(k)+\rho_{h,0}(k)}}K(k,k^{\pr})
\sqrt{{\rho_{0}(k^{\pr})\rho_{h,0}(k^{\pr})
\Delta k^{\pr}\over\rho_{0}(k^{\pr})+\rho_{h,0}(k^{\pr})}}
\ee
Finally, changing the integration variable to $\xi$ in (\ref{funcintdBg})
we have
\bea
Z&=&e^{-\beta F_{\rm min}}
\int\cdots\int\prod_k\left({d\xi(k)\over\pi}\right)
\exp\left\{-\sum_{k,k^{\pr},k^{\dpr}}\xi(k)\left(\delta_{k,k^{\pr}}-M_{k,k^{\pr}}\right)
\left(\delta_{k^{\dpr},k^{\pr}}-M_{k^{\dpr},k^{\pr}}\right)\xi(k^{\dpr})\right\}\nn\\
&&\\
&=&e^{-\beta F_{\rm min}}
\left(\det\left[\delta_{k,k^{\pr}}-K_{k,k^{\pr}}\right]\right)^{-1}\,
\nn
\eea
with
\be
K_{k,k^{\pr}}={\sqrt{\rho_{h,0}(k)}}\cdot M_{k,k^{\pr}}\cdot
{1\over\sqrt{\rho_{h,0}(k^{\pr})}}=\sqrt{{\rho_{0}(k)
\Delta k\over\rho_{0}(k)+\rho_{h,0}(k)}}K(k,k^{\pr})
\sqrt{{\rho_{0}(k^{\pr})
\Delta k^{\pr}\over\rho_{0}(k^{\pr})+\rho_{h,0}(k^{\pr})}}\,.
\ee
The determinant in the above formula evaluated using the 
identities
\be\label{detexp}
\left(\det\left[{\bf{I}}-{\bf{K}}\right]\right)^{-1}=\exp\left\{-{\rm Tr}
\ln({\bf{I}}-{\bf{K}})\right\}=
\exp\left\{\sum_n{1\over n}{\rm Tr}{\bf{K}}^n\right\}
\ee
and taking the $\sum_k\Delta k\to\int dk$ limit yields
\be
Z=e^{-\beta F_{\rm min}+\cS}
\ee
where
\be\label{cS}
\cS=\sum_n{1\over n}\cK^n
\ee
with
\bea\label{cKn}
\cK^n&=&\int_{-\infty}^{\infty}\cdots\int_{-\infty}^{\infty}{dk_1}\cdots{dk_n}\times\\
&&{\rho_0(k_1)\over\rho_0(k_1)+\rho_{h,0}(k_1)}K(k_1,k_2)
{\rho_0(k_2)\over\rho_0(k_2)+\rho_{h,0}(k_2)}\cdots
{\rho_0(k_n)\over\rho_0(k_n)+\rho_{h,0}(k_n)}K(k_n,k_1)\nn\\
&&\nn\\
&=&\int_{-\infty}^{\infty}\cdots\int_{-\infty}^{\infty}{dk_1}\cdots{dk_n}
{1\over1+e^{\beta\epsilon(k_1)}}K(k_1,k_2)
{1\over1+e^{\beta\epsilon(k_2)}}\cdots
{1\over1+e^{\beta\epsilon(k_n)}}K(k_n,k_1)\,.
\nn\eea
(We have to note here, that, although our notation suggests so, for $T\not=0$ $\cS$ is not a correction to the entropy, nevertheless we use this notation, as formally it comes from the 
density of states.) 

Now, similarly to the case of the free Fermi gas we should make a kind of 
"validity test". In this, in addition to the questions discussed there
(application of the free energy functional, and approximations in the evaluation
of the partition function)
one hast to see also, that the $\sum_n{1\over n}\cK^n$ is convergent. 
\begin{enumerate}
\item\emph{The free energy functional.} While in the case of the free Fermi gas the application of
the free energy functional introduced to the free energy density an $O\left((\Delta{k})^2\right)$ error only, in the present case we are faced to an apparently more serious problem, which originates
from the fact, that the   
$J_i$ quantum numbers in (\ref{LiLie}) are either integers or half-integers depending on the parity of the particle number. Changing all $J_i$ quantum numbers from integer to half-odd-integers or 
vice versa shifts all $k_i$ by $\pi/L$ leading to a shift also in the free energy. For general 
$\rho(k)$s this shift can be of $O(1)$ suggesting that the free energy for this system as a function of the $\rho(k)$ is defined with an $O(1)$ accuracy only. Fortunately not this is the case. In Appendix \ref{sec:even-odd} we show, that the above 
used definition of the free energy is accurate enough
for all the $\rho(k)$s contributing to the $Z$ significantly, as the uncertainty coming from the
prescriptions for the $J_i$s is $O(1/\sqrt{L})$, i.e.~it disappears in the $L\to\infty$ limit.  

\item\emph{The approximations in evaluating the $Z$.} 
This group of questions is completely analogous to the questions emerged
in connection to the free Fermi gas, and the answers are similar too: 
in an appropriate cutoff procedure involving the limits 
$L\to\infty$, $\Delta k\to 0$ and $\Lambda\to\infty$ (Appendix \ref{sec:lambdacutoff})
Stirling's formula can be applied in its (\ref{lnomegaro}) form,
$\varsigma(\rho)$ can be taken at 
$\rho_0$ and $\rho_{h,0}$ and also the limits of the 
$\xi$ integrals can be taken to infinity. (The major point in this is that the 
variable $r\propto\xi/\sqrt{L\Delta k}$, i.e.~only a $\sim1/\sqrt{L\Delta k}$
neighborhood of the $\rho_0$ and $\rho_{h,0}$ plays any role.)

\item \emph{The application of the formulas} (\ref{detexp}) (the convergence of (\ref{cS})). The condition for this is that all the eigenvalues of the 
matrix ${\bf{K}}$ are of modulus less than one. In the Appendix \ref{sec:largesteigenvalue} we show that this is true for any $T>0$, and we show also, that the $T\to0$ limit of  
the sum $\sum_n{1\over n}\cK^n$ exists too. 
\end{enumerate}

As we have seen, our method is strictly established in the
$L\to\infty$ limit only, but in a less strict manner we may say, however,
that the $O(1)$ corrections are correctly given also for finite but large enough
size too: although in that case the $\Lambda\to\infty$ limit can not be completed
(as the $L\Delta k\rho(\Lambda)\gg1$ would not hold for very large $\Lambda$),
the contribution of the states above $\Lambda$ would be suppressed anyhow due to the large 
energy. For large but finite system, however, one has to be careful with the
$T\to0$ limit. 
The quantity $\cS$
is an $O(1)$ correction to the thermodynamic potential:
\be
-T\ln Z=F_{\rm min}-T\cS\,. 
\ee
As for $T\to0$ $F_{\rm min}\to E_0$ (with $E_0$ being the ground state energy)
$\lim_{T\to0}\cS$ is an entropy. For infinite $L$ this is the residual
entropy which can be finite, but  
for large but finite $L$ $\lim_{T\to0}\cS$ should be zero as the ground state is 
non degenerated. The resolution of this contradiction is that for finite $L$
our calculation breaks down at temperatures 
where $L\Delta k\rho_{0,h}(0)\sim1$, and below this temperatures the $O(1)$
corrections gradually disappear. 
For this reason the $L\to\infty$ and $L\gg1$
cases should be distinguished!

\goodbreak
\begin{flushleft}
\emph{The case of open ends}
\end{flushleft}
The system, just as in the previously discussed case, is described by the Hamiltonian
(\ref{deltaHam}), but the quantization condition is different: now the ring is not 
closed, and the particles are reflected on the ends. We suppose, that if a particle with 
wavenumber $-k$  arrives at the end at $x=0$, it is reflected to have a wavenumber $k$
while the phase of the wavefunction is shifted by $\varphi_0(k)$:
\be 
e^{-ikx}\longrightarrow e^{ikx+i\varphi_0(k)}\quad\quad(x\sim 0)\,.
\ee
Similarly, a particle of 
wavenumber $k$  arriving at the end at $x=L$ is reflected to have a wavenumber $-k$
while the phase of the wavefunction is shifted by $\varphi_L(k)$:
\be 
e^{ikx}\longrightarrow e^{-ikx+2ikL+i\varphi_L(k)}\quad\quad(x\sim L)\,.
\ee
(If the system is closed by infinitely high potential walls, $\varphi_0(k)=\varphi_L(k)=\pi$,
but with other choices of $\varphi_0(k)$ and $\varphi_L(k)$ different types of ends can be generated.
In Appendix \ref{sec:surfacepotential} we discus a case, when a surface potential having bound states
closes the chain.) The Bethe Ansatz equations of such a system read:
\be\label{LiLiemod}
2Lk_i=2\pi J_i-\varphi_0(k_i)-\varphi_L(k_i)-
\sum_{j(\not=i)}^N \left\{2\atn\left({k_i-k_j\over c}\right)
+ 2\atn\left({k_i+k_j\over c}\right)\right\}
\,.
\ee
Here the $J_i$ numbers are integers, $k_j\not=\pm k_i$ for $j\not=i$, and none of the
$k_j$s equals zero. This means, the real $k_j$ are different positive numbers. Now we 
suppose, the $\varphi_0(k)$ and $\varphi_L(k)$ phases do not generate surface bound states 
(i.e.~states with complex $k$s) and we discus the distribution of the real $k$s (the generalization for the case of surface bound states is straightforward, as it is seen in Appendix \ref{sec:surfacepotential}).  

To proceed we split up the positive $k$ axis into $\Delta k$ intervals, and introduce the densities
of the particles and holes ($\OL\rho(k)$ resp.~$\OL\rho_h(k)$) in the usual manner. Now the integral
equation connecting these quantities is
\bea\label{rorohdBgOE}
\OL\rho(k)+\OL\rho_h(k)&=&\sigma(k)+\sum_{k^{\prime}}\OL{K}\left(k,k^{\pr}\right)\OL\rho(k^{\prime})\Delta k^{\prime}\nonumber\\
&=&\sigma(k)+\int_{0}^{\infty}\OL{K}\left(k,k^{\pr}\right)\OL\rho(k^{\prime})dk^{\prime}
\eea
with 
\be
\sigma(k)={1\over2\pi}\left\{2+{1\over L}{\partial\varphi_0(k)\over\partial k}
+{1\over L}{\partial\varphi_L(k)\over\partial k}-{1\over L}{4c\over c^2+(2k)^2}\right\}\,,
\ee
and
\be
\OL{K}\left(k,k^{\pr}\right)={1\over2\pi}\left({2c\over c^2+\left(k-k^{\pr}\right)^2}+
{2c\over c^2+\left(k+k^{\pr}\right)^2}\right)\,.
\ee
Now the number of states characterized by the same $\OL\rho(k)$ function is
\be
\OL\Omega\left[\OL\rho(k)\right]=
\left.{L\Delta k(\OL\rho(k)+\OL\rho_h(k))-1/2\choose L\Delta k\OL\rho(k)}\right|_{k=0}
\prod_{k\not=0}\omega(\OL\rho(k))\,.
\ee
Here $k=0$ refers to the interval beginning at the origin, and 
\be
\omega(\OL\rho(k))={L\Delta k(\OL\rho(k)+\OL\rho_h(k))\choose L\Delta k\OL\rho(k)}\,,
\ee
just as previously. The contribution of the $k=0$ interval differs from those of the others as the 
$k=0$ wavenumber (what is right on the edge of the interval) is not allowed. (In a 
$2\Delta k$ interval containing the origin in the middle there are 
$2L\Delta k(\OL\rho(k=0)+\OL\rho_h(k=0))$ $k$ values, out of this 
$L\Delta k(\OL\rho(k)+\OL\rho_h(k))-1/2$ is positive, 
thus this is the number of choosable $k$s in the $k=0$ interval.)
Applying Stirling's formula we arrive at 
\be
\OL\Omega\left[\OL\rho(k)\right]=
\sqrt{\left.\left({\OL\rho_h(k)\over\OL\rho(k)+\OL\rho_h(k)}\right)\right\vert_{k=0}}
\prod_{k}\omega(\OL\rho(k))\,.
\ee

The free energy functional is
\be\label{fefdBgOE}
\OL F\left[\OL\rho(k)\right]=\sum_k\left(
Le(k)\OL\rho(k)\Delta k-T\ln\omega(\OL\rho(k))\right)-
{T\over2}\ln\left.{\OL\rho_h(k)\over\OL\rho(k)+\OL\rho_h(k)}\right|_{k=0}\,.
\ee 
The minimization of the free energy leads to the condition
\be
{\OL\rho_0(k)\over \OL\rho_{h,0}(k)}=e^{-\beta\epsilon(k)}\,,
\ee
with $\epsilon(k)$ given by (\ref{epsilon}) (just as in the PBC case), and it yields a minimal value
\bea
\OL F_{\rm min}&=&-{L\over\beta}\sum\ln\left(1+e^{-\beta\epsilon(k)}\right)\sigma(k)\Delta k
+{T\over2}\ln\left(1+e^{-\beta\epsilon(0)}\right)\nonumber\\
&=&-{L\over\beta}\int_{0}^{\infty}
\ln\left(1+e^{-\beta\epsilon(k)}\right)\sigma(k)dk
+{T\over2}\ln\left(1+e^{-\beta\epsilon(0)}\right)\,.
\eea
Finally the evaluation of the functional integral leads to
\be
\OL Z=e^{-\beta \OL F_{\rm min}+\OL{\cS}}
\ee
with
\be
\OL{\cS}=\sum_n{1\over n}\OL \cK^n
\ee
where
\bea
\OL \cK^n&=&\int_{0}^{\infty}\cdots\int_{0}^{\infty}{dk_1}\cdots{dk_n}\times\\
&&{\OL\rho_0(k_1)\over\OL\rho_0(k_1)+\OL\rho_{h,0}(k_1)}\OL K(k_1,k_2)
{\OL\rho_0(k_2)\over\OL\rho_0(k_2)+\OL\rho_{h,0}(k_2)}\cdots
{\OL\rho_0(k_n)\over\OL\rho_0(k_n)+\OL\rho_{h,0}(k_n)}\OL K(k_n,k_1)\nn\\
&&\nn\\
&=&\int_{0}^{\infty}\cdots\int_{0}^{\infty}{dk_1}\cdots{dk_n}
{1\over1+e^{\beta\epsilon(k_1)}}\OL K(k_1,k_2)
{1\over1+e^{\beta\epsilon(k_2)}}\cdots
{1\over1+e^{\beta\epsilon(k_n)}}\OL K(k_n,k_1)\,.
\nn\eea
Using the actual form of $\sigma(k)$ and $\OL K(k,k^{\pr})$ it is easy to see,
that
\be
\OL F_{\rm min}=F_{\rm min}+\Delta F+\phi_0+\phi_L\,,
\ee
with
\be\label{DeltaF}
\Delta F={T\over2}\ln\left(1+e^{-\beta\epsilon(0)}\right)+
{T\over2\pi}\int_0^{\infty}{4c\over c^2+(2k)^2}\ln\left(1+e^{-\beta\epsilon(k)}\right)dk\,,
\ee\be
\phi_{0/L}=-{T\over2\pi}\int_0^{\infty}{\partial\varphi_{0/L}(k)\over\partial k}
\ln\left(1+e^{-\beta\epsilon(k)}\right)dk\,.
\ee
We note here the following. The macroscopic part of the thermodynamic potential $F_{\rm min}$ is independent
of the boundary condition as it should be. $\Delta F$ is a consequence 
of the openness of the chain, but the nature of the surfaces is reflected
in $\phi_0$ resp.~$\phi_L$ only. These three $O(1)$ corrections directly connected to the surfaces are given already by the usual thermodynamic treatment. The contribution of the saddle point fluctuations depends on the boundary conditions in its structure. To see the difference between the $\cS$ valid for the PBC and $\OL\cS$ applying for the open ends we introduce the function 
\be\label{widetildeK}
\widetilde{K}(k,k^{\pr})=\sum_{n=1}^{\infty}{1\over n}\widetilde{K}^n(k,k^{\pr})
\ee
with
\bea
&&\widetilde{K}^n(k,k^{\pr})=\int_{-\infty}^{\infty}\cdots\int_{-\infty}^{\infty}{dk_1}\cdots{dk_{n-1}}\times\\
&&\sqrt{1\over1+e^{\beta\epsilon(k)}}K(k,k_1)
{1\over1+e^{\beta\epsilon(k_1)}}
K(k_1,k_2){1\over1+e^{\beta\epsilon(k_2)}}
\cdots
{1\over1+e^{\beta\epsilon(k_{n-1})}}K(k_{n-1},k^{\pr})\sqrt{1\over1+e^{\beta\epsilon(k^{\pr})}}\,.\nn
\eea
Using the form of $K(k,k^{\pr})$ and $\OL K(k,k^{\pr})$ it is not hard to see, that with this notation
\be
\cK^n=\int_{-\infty}^{\infty}\widetilde{K}^n(k,k)dk,\ \ \ {\rm and}\ \ \ \OL\cK^n
=\int_{0}^{\infty}\widetilde{K}^n(k,k)dk+\int_{0}^{\infty}\widetilde{K}^n(k,-k)dk\,,
\ee
thus
\be\label{difference}
\cS=\int_{-\infty}^{\infty}\widetilde{K}(k,k)dk,\ \ \ {\rm while}\ \ \ \OL\cS
=\int_{0}^{\infty}\widetilde{K}(k,k)dk+\int_{0}^{\infty}\widetilde{K}(k,-k)dk\,.
\ee
We note that the considerations presented in Appendix \ref{sec:largesteigenvalue} concerning the convergence of $\cS$ hold for $\OL\cS$ too, i.e.~also $\OL\cS$ together with its $T\to0$ limit exists.

\section{Generalization}
\label{sec:Generalization}

The repulsive $\delta$ Bose gas is the simplest BA system in the sense, that the particles have no internal structure and do not form bound states allowing to describe the system by one set of real
parameters (the wavenumbers). In most of the BA systems, however, for the thermodynamic description one has to introduce many (in most of the cases \emph{infinitely} many) sets of rapidities. These can be of different type. For example in the case of Heisenberg chain these 
variables are the centers of the strings of different length \cite{TakaSuzu}, 
for the repulsive $\delta$ Fermi gas one set gives the wavenumbers, and the others are strings 
connected to the spin state of the system,
while in case of the Hubbard model there are three type of rapidities: the real wavenumbers, the centers of the strings connected to the spins and the centers of the strings connected to the bound pairs \cite{Taka}. (Here we take granted, that the string hypothesis works, as all in the known cases, where
independent check is possible, it gives the correct result \cite{DeEssGohKluKorKus}.) These systems are much more complicated than the $\delta$ Bose gas, nevertheless it seems, that our calculation is generalizable for a larger class of them. Now we outline this procedure. 

For the sake of simplicity we denote all of the rapidities by $k$, this will cause no confusion.
To each set of rapidities particle and hole densities ($\rho^{(n)}(k)$ resp.~$\rho_h^{(n)}(k)$) can be defined. Consider a system in which these satisfy a set of integral equations of the type:
\be\label{BAgen}
\rho^{(n)}(k)+\rho_h^{(n)}(k)=\sigma_n(k)+\sum_m\int K_{n,m}(k,k^{\pr})\rho^{(m)}(k^{\pr})dk^{\pr} \,.
\ee
The number of states described by the same set of densities is now supposed to be
\be\label{nosgen}
\prod_n\Omega\left[\rho^{(n)}(k)\right]
\ee
and the free energy functional to be minimized is of the form
\be
\textbf{F}\left[\rho^{(n)}(k)\right]=L\sum_n\int \left(e_n(k)\rho^{(n)}(k)-Ts\left(\rho^{(n)}(k)\right)\right)dk
\ee
where $e_n(k)$ is the energy of an object of type $n$ and rapidity $k$.
The minimization leads to the condition
\be 
{\rho_0^{(n)}(k)\over\rho_{h,0}^{(n)}(k)}=e^{-\beta\epsilon_n(k)}
\ee
with 
\be
\epsilon_n(k)=e_n(k)-T\sum_m\int
\ln\left(1+e^{-\beta\epsilon_m(k^{\prime})}\right)K_{m,n}(k^{\pr},k)dk^{\prime}\,.
\ee
The minimal value of the free energy functional is
\be
\textbf{F}_{\rm min}=-TL\sum_n\int\ln\left(1+e^{-\beta\epsilon_n(k)}\right)\sigma_n(k)dk\,.
\ee
 
The contribution of the states near to the one minimizing the free energy functional
can be calculated by the functional integral method described in the previous section.   
Through a very straightforward calculation one finds, that the contribution of these states  to the 
free energy given as the logarithm of the partition function is
\be\label{TDeltaSgen}
\Delta\textbf{F}=-T\Delta\textbf{S}\,,
\ee
with
\be
\Delta\textbf{S}=\sum_n{1\over n}K^n\,,
\ee
where now   
\bea
K^n=&&\sum_{m_1}\cdots\sum_{m_n}\int\cdots\int{dk_1}\cdots{dk_n}\times\\
&&\nn\\
&&{1\over1+e^{\beta\epsilon_{m_1}(k_1)}}K_{m_1m_2}(k_1,k_2)
{1\over1+e^{\beta\epsilon_{m_2}(k_2)}}\cdots
{1\over1+e^{\beta\epsilon_{m_n}(k_n)}}K_{m_nm_1}(k_n,k_1)\,.
\nn\eea  

The above calculation is a formal generalization of the procedure applied for the $\delta$ Bose gas,
and it involves the same kinds of approximations too. For this in any case one should check, if the
conditions are met. Here we emphasize one point: Stirling's formula is applicable only if for all
$\rho^{(n)}(k)$ giving significant contribution in the integral $L\rho^{(n)}(k)\Delta k\gg1$ and $L\rho_h^{(n)}(k)\Delta k\gg1$. This can be made true taking the $L\to\infty$ limit only in the case, if the 
$\rho_0^{(n)}(k)>0$ and $\rho_{h,0}^{(n)}(k)>0$ for all $n$ and $k$.
 
Our result has a meaning only if the sum defining $\Delta\textbf{S}$ converges. In any known system
the kernel $K_{mm^{\pr}}(k,k^{\pr})$ is rather complicated, and checking the convergence may encounter difficulties. In some simple cases, however, this check is doable: For example in the case of the Heisenberg chain for $T=0$ the density of 1-strings (real rapidities) remains finite only, all other densities
disappear, and the procedure of Appendix \ref{sec:largesteigenvalue} can be applied.
This gives the result that in the case of finite magnetic field the sum is convergent for the complete antiferromagnetic region 
(the anisotropy $\varrho>-1$ in the Hamiltonian (\ref{XXZHam})), but it is not convergent 
in the critical region ($1\geq\varrho>-1$) for 
zero field. This renders it likely, that the sum is convergent for finite temperature too   
for finite field for any $\varrho$, or even for zero field if $\varrho>1$, but does not support any guess for finite temperature and no field if $1\geq\varrho>-1$.

Even if the $\Delta\textbf{S}$ exists, to decide if (\ref{TDeltaSgen}) is correct for a given 
system \emph{may need further considerations}.  
In deriving (\ref{TDeltaSgen})
we assumed that only (\ref{BAgen}) constrains the densities and that in principle all of the 
states are so described, i.e.~(\ref{nosgen}) gives correctly the number of states described by the same density. These assumptions, however, may not be true even in the best known cases. Two kinds of problems may arise:
\begin{enumerate}
	\item\label{one} In addition to (\ref{BAgen}) there are \emph{other constrains} on the densities too. 
	For example in systems of 1/2 spins the 
	total number 
	of the rapidities must not exceed the half of the number of sites/particles. (This expresses
	the fact, that the BA equations describe the states of total spin $S^z\geq0$ only, the $S^z<0$ 
	states are obtained by reflection) 
	 
	\item Not all of the states are automatically described by the possible densities. 
	For example for SU(2) systems the BA equations describe the states of highest weight only, 
	the others are obtained 
	from these by further manipulations. It is also possible, that in the system there are more 
	vaqua, and more sets of excitations, and this multiplicity is not taken into account
	neither in the usual, nor in the present description. 
\end{enumerate}
In the following we discuss these problems briefly in case of the Heisenberg chain.

The XXZ Heisenberg chain of length $N$ in magnetic field $h$ is described by the Hamiltonian
\be\label{XXZHam}
\hat{H}_{XXZ}=\sum_{i}^N \left(S_i^x S_{i+1}^x+S_i^y S_{i+1}^y+\varrho S_i^z S_{i+1}^z-hS_i^z\right)\,.
\ee
This model is Bethe Ansatz diagonalisable, but the BA equations give certain classes of the 
eigenstates only, and the others should be constructed by further manipulations.   
As the basic properties of the model are concerned four different case should be distinguished, but from the thermodynamic point of view the following three \emph{antiferromagnetic} cases are important.

\emph{The $-1<\varrho<1$ planar case.} In this case the BA equations give the $S^z=\sum_iS_i^z\geq0$
states only: in all of the solutions the total number of turned down spins (spin-waves) is less than or equal to the half of
the number of sites $N/2$. As it has been mentioned in point \ref{one}.~above this imposes a constrain on the densities as they should satisfy the sum rule 
\be\label{addcontstrain}
{1\over2}\geq\sum_{(n)}l_{(n)}\int\rho^{(n)}(k)dk\,,
\ee
with $l_{(n)}$ being the length of the string-type labeled by $(n)$. Those densities not obeying
this constrain are nonphysical, this can be manifested in negative $\rho_h(k)$'s generated by (\ref{BAgen}). If in (\ref{addcontstrain}) the equality holds, than $S^z=0$, if the $>$ sign is valid, than $S^z>0$, and the $S^z<0$ states are constructed by reflection of the $S^z>0$ ones. 
This means, that calculating the partition function the contributions of the 
$S^z>0$ states have to be taken into account with weights 
\be
1+\exp\{-2\beta S^zh\}\,,
\ee
i.e.~all $S^z>0$ states have additional contributions to the free energy
\be
-T\ln\left(1+\exp\{-2\beta S^zh\}\right)\,,
\ee
but no such contributions exist for the $S^z=0$ states.
For finite $h$ the magnetization is macroscopic, $S^z\propto L$, thus if $L\to\infty$, the above contribution to the free energy
disappears, but for $h=0$ it remains finite $-T\ln2$. On the other hand 
in finite magnetic field all the densities contributing significantly to the functional integral obey the sum rule (the equilibrium densities satisfy (\ref{addcontstrain}) with the sign $>$), thus the correction (\ref{TDeltaSgen}) is certainly correct, provided it is convergent. This is also true for zero magnetic field, although the situation in that case is somewhat different. In zero field the equilibrium densities describe states with zero magnetization (i.e.~they satisfy (\ref{addcontstrain}) with the sign $=$), and  
the functional integral also involves nonphysical densities (which violate the sum rule).
Taking into account the symmetry of the functional integral it is clear, however, that the contribution 
of the nonphysical densities is the same as the contribution of the densities describing
$S^z>0$ (physical) states, i.e.~the functional integral takes into account the $S>0$ states with a weight of two -- just as it should be.

\emph{The $\varrho=1$ isotropic chain.} In this case in addition to the above constrain an other
problem arises. The model has an SU(2) symmetry, and the BA equations describe the highest weight ($S^2=S^z(S^z+1)$) states only, the other states of the 
same spin length are obtained by means of the $S^-$ operator. For this, if $h\not=0$, the  contribution to the partition function of one highest weight state should be weighted by
\be
{1-\exp\{-\beta h(2S^z+1)\}\over1-\exp\{-\beta h\}}
\ee
to get the contribution of the complete multiplet, i.e.~to the free energy functional a term
\be
-T\ln\left({1-\exp\{-\beta h(2S^z+1)\}\over1-\exp\{-\beta h\}}\right)
\ee
should be added. In the thermodynamic limit $\exp\{-\beta h(2S^z+1)\}$ disappears, and we find an additional $O(1)$ correction
\be
T\ln\left(1-\exp\{-\beta h\}\right)
\ee
to the free energy. Note, that this is divergent if $h\to0$ indicating, that the corresponding 
correction in the $h=0$ case is of a different order of magnitude. (This divergence is a 
consequence of the thermodynamic limit, for finite $N$ this term would behave as  
$-T\ln(2S^z(h)+1)$.)
   
In the case of zero magnetic field all members of a spin multiplet are of the same energy, thus
in order to get the contribution of a complete multiplet characterized by a spin length $\cL$
(this $\cL$ being equal to the $S^z$ of the highest weight member)
a correction 
\be 
-T\ln(2\cL+1)
\ee
should be added to the free energy functional. Although for the equilibrium density this correction is zero, we show in Appendix \ref{sec:ln}, that it can be very large when deviating from the equilibrium 
density, thus its treatment requires a much more subtle procedure, what is beyond the grasp of 
the present work. We also give an estimation according to which this term is $O(\ln N)$, so we have to conclude that for the isotropic chain in zero field the leading correction to the macroscopic free energy may come from this term, not from the saddle point fluctuations.

\emph{The $\varrho>1$ "easy axis" anisotropic chain.} In this case -- similarly to the $\varrho<1$ case -- the solutions of the BA equations
give all of the $S^z\geq0$ states, and the $S^z<0$ states are obtained by reflecting the $S^z>0$ ones. In this sense the $\varrho>1$ and $\varrho<1$ are in close analogy and the conclusions concerning the $O(1)$ corrections made for $\varrho<1$
hold also for $\varrho>1$. Nevertheless in this case there is an additional problem. In this region
the vacuum is twofold degenerated, this is manifested in the fact, that from the ground-state density
of rapidities two different solution of the BA equations can be reconstructed. Also the low energy
excited states can be grouped into two sets as being the excitations above one or the other vacuum
\cite{ViroWoy}, but it has not been studied yet, if such a degeneracy exists also in the thermodynamically important highly excited states. It is also not claerd yet, if our method takes this kind of degeneracy automatically into account, although this can be important, as it is expected to give an $O(1)$ correction too, which behaves like $-T\ln2$ as $T\to0$.

\section{Summary}\label{sec:summary}  

In the present work based on the method deviced by Yang and Yang \cite{CNYCPY} we developed a functional integral method to calculate $O(1)$ corrections
to the free energy of macroscopic BA integrable systems. In the Yang and Yang method the free energy
of a system is written up as a functional of the momentum density, and this functional is minimized in order to find the actual value of the free energy. In terms of the grand canonical
partition function the equilibrium density of momenta (at which the free energy is minimal) defines the states entering into the partition function with highest weights. 
The basic point of our calculation is
that in evaluating the grand canonical partition function after the minimization of the free 
energy functional, (what actually gives the macroscopic part,) the contribution of the states near to
the equilibrium (saddle point fluctuations) can be calculated by a Gaussian integral. To define this integral properly one needs to calculate the entropy entering into the free energy functional up to next to leading order. 

In addition to the technical problems the calculation of non macroscopic corrections to the macroscopic free energy  rises  
some conceptional questions too. The Yang and Yang method has been developed to pick up the 
leading contribution only, thus in calculating further terms one has to see, that this refinement is meaningful, the method is accurate enough to calculate the next to leading contributions too.
This involves two kinds of problems. The first is if it is possible at all to define an accurate enough free energy density in terms of the momentum density. (Questions of this type are discussed in Appendices \ref{sec:errorinfeef} and \ref{sec:even-odd}.) The other kind of problem is connected to the accuracy by which the macroscopic part itself is calculated: in systems, in which the number of available momenta is infinite, a cutoff procedure must be introduced even to calculate the macroscopic free energy. (The problem of this cutoff procedure and its resolution in the thermodynamic limit is discussed in Appendix \ref{sec:lambdacutoff}.) 

In order to avoid difficulties of special
models we write up and discuss the method using the free Fermi gas. Our calculation reproduces the exact result, but we see that our method for the above mentioned implicit cutoff procedure is established in a strict mathematical sense 
in the thermodynamic limit only.

Next the method is generalized for the repulsive $\delta$ Bose gas with PBS. The structure of the 
BA equations for the rapidity densities of this system is (\ref{rorohdBg})
\be\label{rorohdBg2}
\rho(k)+\rho_h(k)
=\sigma(k)+\int_{-\infty}^{\infty}
K(k,k^{\pr})\rho(k^{\prime})dk^{\prime}
\ee
with
\be
\sigma(k)={1\over2\pi}\,,\quad\quad\quad K\left(k,k^{\prime}\right)=
{1\over2\pi}\,{2c\over c^2+\left(k-k^{\prime}\right)^2}\,,
\ee
and the energy associated with a particle is 
\be 
e(k)=k^2-\mu\,.
\ee
The $O(1)$ correction to the 
free energy we find is of the form $-T\cS$ with $\cS$ given in the form of an 
infinite sum (\ref{cS}-\ref{cKn}) or equivalently in the form (\ref{widetildeK})-(\ref{difference})
\be\label{differenceeleje}
\cS=\int_{-\infty}^{\infty}\widetilde{K}(k,k)dk,
\ee
where
\be\label{widetildeKujra}
\widetilde{K}(k,k^{\pr})=\sum_{n=1}^{\infty}{1\over n}\widetilde{K}^n(k,k^{\pr})
\ee
with
\bea
&&\widetilde{K}^n(k,k^{\pr})=\int_{-\infty}^{\infty}\cdots\int_{-\infty}^{\infty}{dk_1}\cdots{dk_{n-1}}\times\\
&&\sqrt{1\over1+e^{\beta\epsilon(k)}}K(k,k_1)
{1\over1+e^{\beta\epsilon(k_1)}}
K(k_1,k_2){1\over1+e^{\beta\epsilon(k_2)}}
\cdots
{1\over1+e^{\beta\epsilon(k_{n-1})}}K(k_{n-1},k^{\pr})\sqrt{1\over1+e^{\beta\epsilon(k^{\pr})}}\nn
\eea
and the dressed energy $\epsilon(k)$ entering this formula being the usual one given 
by the equation (\ref{epsilon})
\be\label{epsilon2}
\epsilon(k)=e(k)-{T}\int_{-\infty}^{\infty}
K(k^{\prime},k)\ln\left(1+
e^{-\beta\epsilon(k^{\prime})}\right)dk^{\prime}\,.\nn
\ee

We have shown also, that $\cS$ and its $T\to0$ limit exist and are finite. In the proof it is used
that the particle density is a finite value fixed by the chemical potential $\mu$. The "entropy"
$\lim_{T\to0}\cS$ is the residual entropy not equal to the entropy of the (unique) ground state what
would be zero.
This is a consequence of the fact that the $T\to0$ limit is taken after the $L\to\infty$ limit,
and these limits do not commute.

We also calculated the contribution of the saddle point fluctuations to the free energy of
open end systems. We have found that it is slightly different in structure (\ref{difference}). It is $-T\OL{\cS}$
with   
\be\label{correction2}
\OL\cS
=\int_{0}^{\infty}\widetilde{K}(k,k)dk+\int_{0}^{\infty}\widetilde{K}(k,-k)dk\,.
\ee

In field theory systems of massive relativistic particles are important in which the particle number is not regulated by a chemical potential. A system of this type is described by the Lee-Yang model \cite{LeeYang} 
in which $e(k)=M\ch(k)$
and the kernel $K$ is negative. This system behaves somewhat differently 
than the $\delta$ gas. There are strong arguments supporting the claim, that for PBC no $O(1)$ corrections shoud be present. Contrary to this our method gives 
a $\cS$ of the type (\ref{differenceeleje}) for this case. 
For large $T$ $\exp\{-\beta\epsilon(k)\}$ is a constant for $e(k)<T$
and is zero above. In this case $\cS\propto-\ln(T/M)$. If $T\to0$, then $\epsilon(k)\to e(k)$,
$\exp\{\beta\epsilon(k)\}$ diverges and $\cS$ disappears. For the open end case we get a correction of the type (\ref{correction2}) what also behaves differently as expected: due to the first term
it diverges in the $T\to\infty$ limit. We note, that in a recent work \cite{DoFiChTa} based on a different type of calculation it has been proposed, that the $O(1)$ correction in the open end case sholud be of the
type of the second term in (\ref{correction2}).

One has to note here the following. It is possible to define some field theoretical models 
as certain limits of lattice models, for example the scaling limits of the Hubbard or the 
Heisenberg models are closely related to the $SU(2)$ chiral invariant Gross-Neveu model
\cite{WoFo1,WoFo2,HaRaWo}. Calculating the $O(1)$ corrections in the lattice models and taking the scaling limit afterward we expect $\lim_{T\to0}\cS$ involve the residual entropy of the vacuum too. 

We also give a generalization of our result to other Bethe Ansatz systems. In a large class of models the densities satisfy equations of the type (\ref{BAgen})
\be
\rho^{(n)}(k)+\rho_h^{(n)}(k)=\sigma_n(k)+\sum_m\int K_{n,m}(k,k^{\pr})\rho^{(m)}(k^{\pr})dk^{\pr}\,.
\ee
For these systems we find, that the contribution of the saddle point fluctuations to the free energy is
$-T\Delta\textbf{S}$ (\ref{TDeltaSgen})  with
\be
\Delta\textbf{S}=\sum_n{1\over n}K^n\,,
\ee
where now   
\bea
K^n=&&\sum_{m_1}\cdots\sum_{m_n}\int_{-\infty}^{\infty}\cdots\int_{-\infty}^{\infty}{dk_1}\cdots{dk_n}\times\\
&&\nn\\
&&{1\over1+e^{\beta\epsilon_{m_1}(k_1)}}K_{m_1m_2}(k_1,k_2)
{1\over1+e^{\beta\epsilon_{m_2}(k_2)}}\cdots
{1\over1+e^{\beta\epsilon_{m_n}(k_n)}}K_{m_nm_1}(k_n,k_1)\,.
\nn\eea
In this formula the energies $\epsilon_{n}(k)$ are connected to the $e_{n}(k)$ bare ones 
by the equations 
\be
\epsilon_n(k)=e_n(k)-T\sum_m\int
\ln\left(1+e^{-\beta\epsilon_m(k^{\prime})}\right)K_{m,n}(k^{\pr},k)dk^{\prime}\,.
\ee  
As this formula is a general one, its convergence should be checked in any special case. 
We also point out, that in the special cases additional problems requiring further
considerations may arise, as we illustrate 
on the example of the XXZ Heisenberg chain.

\acknowledgements

I am grateful to Prof.~L.~Palla and Drs.~Z.~Bajnok and G.~Tak\'acs for the valuable discussions.
Work has been supported by OTKA under grant Nr.~T 043330

\appendix
\section{}\label{sec:errorinfeef}
In this Appendix we want to examine the approximation (\ref{summationapproximation}) reading as 
\be\label{summationapproximation2}
	\sum_{\{k_j\}}\exp\left\{-\beta\sum_{k_j}\left(k_j^2-\mu\right)\right\}
	\,\longrightarrow\,   
	\exp\left\{-\beta\left(\overline{k}^2-\mu\right)L\rho(k)\Delta{k}+\ln\omega(\rho(k))\right\}
	\ee
in more detail. Suppose that $\overline{k}$ is the mean value of the $k$s in the $\Delta{k}$ interval,
and that the $\Delta{k}$ is small enough, so we may linearize around $\overline{k}$. Thus to get the 
left hand side of the above formula we have to calculate 
\be 
\exp\left\{-\beta\left(\overline{k}^2-\mu\right)L\rho(k)\Delta{k}\right\}
\sum_{\{n_i\}}\exp\left\{2q\sum_{i=1}^m n_i\right\}\,,
\ee
where $q=-\beta\overline{k}\Delta{k}/N$, the numbers $n_i$ are integers or half-integers 
($n_i=(N-1)/2\modegy$) satisfying
$-(N-1)/2\leq n_1<\ldots<n_i<n_{i+1}<\ldots<n_m\leq(N-1)/2$ with $N=L(\rho(k)+\rho_0(k))\Delta{k}$
and $m=L\rho(k)\Delta{k}$, and the $\sum_{\{n_i\}}$ extends over all possible $n_i$ sets. This gives
\be 
\exp\left\{-\beta\left(\overline{k}^2-\mu\right)L\rho(k)\Delta{k}\right\}
\prod_{i=1}^m{\sh((N-i+1)q)\over\sh(iq)}\,,
\ee
what for small enough $\Delta{k}$ yields
\be 
\exp\left\{-\beta\left(\overline{k}^2-\mu\right)L\rho(k)\Delta{k}+\ln\omega(\rho(k))+
{q^2\over6}\sum_{i=1}^m\left((N-i+1)^2-i^2\right)\right\}\,.
\ee
Evaluating the sum and inserting the value of $N$ and $m$ we arrive at 
\be 
\exp\left\{-\beta\left(\overline{k}^2-\mu-{\beta\overline{k}^2(\Delta{k})^2\over6}{\rho_h(k)\over
\rho(k)+\rho_h(k)} \right)L\rho(k)\Delta{k}+\ln\omega(\rho(k))\right\}\,,
\ee
It is not hard to see, that taking into account the quadratic nature of the spectrum
would lead also to corrections not larger than $O((\Delta{k})^2)$, i.e.~the correction to the free energy 
density neglected in (\ref{summationapproximation}) ((\ref{summationapproximation2})) is
indeed small enough to disappear in the $\sum(\ldots)\Delta{k}\longrightarrow\int(\ldots)dk$ limit.

\section{}\label{sec:even-odd}

Here we show, that the parity prescription for the parameters $J_i$ ($J_i=(N+1)/2\modegy$) does not destroy the accuracy of the free energy functional. Let us consider two systems, one described by the system of equations 
\be\label{LiLie2}
Lk_i=2\pi J_i-\sum_j^N 2\atn\left({k_i-k_j\over c}\right)
\ee
with $J_i$ being integer, the other with $J_i$ being half-odd-integer independently from the
particle number.  Let us distinguish between the parameters of the two systems by primes
($k_i^{\pr}$, $J_i^{\pr}$) and double primes ($k_i^{\dpr}$, $J_i^{\dpr}$). (The real system 
is between this two: a primed solution is to be taken if the number of particles is odd, and 
a doubly primed one applies for an even number of particles.) Obviously the two kinds 
of solutions are in one-to-one correspondence: we consider a primed and a doubly primed solution
one pair, if $J_i^{\dpr}=J_i^{\pr}+{1\over2}$ for all $i$. Due to the 'Galilei invariance' of 
(\ref{LiLie2}) the wavenumbers of the pairs are closely related: $k_i^{\dpr}=k_i^{\pr}+\pi/L$. For this we can describe the pairs
by the same $\rho(k)$ and $\rho_h(k)$ if the $\Delta{k^{\pr}}$ and $\Delta{k^{\dpr}}$ intervals are the 
same just shifted by $\pi/L$ relative to each other. Let us denote the free energy associated 
to a $\rho(k)$ in the two systems by $F^{\pr}[\rho(k)]$ and $F^{\dpr}[\rho(k)]$, respectively.
Their difference is
\be 
\Delta{F[\rho(k)]}=  F^{\dpr}[\rho(k)]-F^{\pr}[\rho(k)]\simeq\sum_k2\overline{k}\pi\rho(k)\Delta{k}\,,
\ee
where on the right hand site we dropped the primes. For a general $\rho(k)$ this can be of $O(1)$,
but for those densities which contribute to $Z$ it is much smaller: near to the equilibrium $\rho(k)=\rho_0(k)+r(k)$, and as $\rho_0$ is an even function of $k$,
\be\label{error}
\Delta{F[\rho(k)]}\simeq\sum_{k>0}2\overline{k}\pi\left(r(k)-r(-k)\right)\Delta{k}\,.
\ee
As in the functional integral the major contribution comes from the region $r\propto1/\sqrt{L}$,
for densities important in calculating the saddle point contributions
$\Delta{F[\rho(k)]}\propto1/\sqrt{L}$ is a good estimation. This shows, that the difference in the free energies of the primed and the doubly primed systems disappears as 
$L\to\infty$. As, however, (\ref{error}) is also an estimate for the error made if the parity of the numbers $J_i$ is not chosen properly we may conclude that the prescription for the $J_i$s ($J_i=(N-1)/2\modegy$) does not influence the $O(1)$ corrections.

\section{}\label{sec:largesteigenvalue}
In this Appendix we show, that all the eigenvalues $\kappa$
of the matrix $\bf{K}$ of (\ref{detexp}) in the case of the $\delta$ Bose gas have 
a modulus less than unity. To do this we use the formula
\be
{\rm max}\ln|\kappa|=\lim_{n\to\infty}{1\over n}\ln|{\rm Tr}{\bf{K}}^n|\,.
\ee
As ${\rm Tr}{\bf{K}}^n$ is definitely positive for all $n$ 
the eigenvalue of largest modulus is positive, i.e. 
\be
{\rm max}\ln|\kappa|=
\ln\kappa_{\rm max}
\ee
As we apply our formulas after the $\sum_k\Delta k\to\int dk$ limit is taken,
we may use ${\rm Tr}{\bf{K}}^n=\cK^n$, that is
\be
\ln\kappa_{\rm max}=\lim_{n\to\infty}{1\over n}\ln \cK^n\,.
\ee
For $T>0$, using the relations
\be
\int^{\infty}_{-\infty}\cdots\int^{\infty}_{-\infty}{dk_1}\cdots{dk_{(n-1)}}
K(k,k_1)\cdots K(k_{(n-1)},k^{\prime})={1\over2\pi}{2nc\over(nc)^2+(k-k^{\prime})^2}\,,
\ee
and 
\be
\int\rho_0(k)=N/L\,,
\ee
$\cK^n$ is overestimated by the formula
\be
\cK^n\leq\left({\rho_0(k)\over\rho_0(k)+\rho_{h,0}(k)}\right)_{\rm max}^{n-1}
{1\over\pi nc}{N/L\over (\rho_0(k)+\rho_{h,0}(k))_{\rm min}}
\ee
This yields
\be
\kappa_{\rm max}\leq\left({\rho_0(k)\over\rho_0(k)+\rho_{h,0}(k)}\right)_{\rm max}\,,
\ee
what for $T>0$ is indeed less than one. 
This proves the applicability of the 
formulas (\ref{detexp}) and the convergence of the series $\sum_n{1\over n}\cK^n$ (and also proves,
that the Neumann series of Eg.(\ref{rho0}) converges).

By a slight modification of the above estimations one can also prove, that
the 
\be
\lim_{T\to0}\sum_n{1\over n}\cK^n
\ee
also exists. For $T=0$
\be
{\rho_0(k)\over\rho_0(k)+\rho_{h,0}(k)}=\cases{1&if $|k|\leq k_F$;\cr
0&otherwise,\cr}
\ee
where $k_F$ is a finite wavenumber, under which all, above which none of the states 
are filled. Observing, that
\bea
\int^{k_F}_{-k_F}\cdots\int^{k_F}_{-k_F}{dk_1}\cdots{dk_{(n-1)}}
K(k,k_1)\cdots K(k_{(n-1)},k)&\leq&\nn\\
\int^{\infty}_{-\infty}\cdots\int^{\infty}_{-\infty}{dk_1}\cdots{dk_{(n-1)}}
K(k,k_1)\cdots K(k_{(n-1)},k)&=&{1\over\pi nc}
\eea
we see, that 
\be
\lim_{T\to0}\sum_n{1\over n}\cK^n\leq {2k_F\over\pi c}\sum_n{1\over n^2}={2k_F\pi\over6c}\,.
\ee

We note, that if $c\to0$, i.e.~if the kernel is a $\delta$-function,
both for $T>0$ and $T\to0$ our estimation blows up.

\section{}\label{sec:lambdacutoff}

In this appendix we discuss the questions connected to the cutoff procedure involved in the evaluation of the free energy. First we notice, that this problem is rather a problem of the 
accuracy of the macroscopic part of the free energy. To see this consider the partition function $Z$ of a system. The macroscopic part of the free energy is defined as 
\be
F_{\rm mac}=Lf\,,\quad\quad\mbox{where}\quad\quad f=-T\lim_{L\to\infty}{1\over L}\ln Z\,. 
\ee
The next to leading correction to the macroscopic part is $O(L^0)$ if 
\be
(\cS=)
\lim_{L\to\infty}\ln\left(Ze^{\beta F}\right)
\ee
is finite but zero. If so, the free energy defined through the logarithm of the partition function
behaves for large enough $L$ as 
\be
F=-T\ln Z=Lf-T\cS\,.
\ee
In our case in order to apply Stirling's formula
we have to introduce the cutoff $\Lambda$ in the $k$ space, and for the same reason
we have to calculate sums instead of integrals. This way the partition function we obtain is of the form
\be
Z_L(\Lambda,\Delta k)=e^{-\beta F_{\rm min}(L,\Lambda,\Delta k)+\Delta S(\Lambda,\Delta k)}
=e^{-\beta Lf_{\rm min}(\Lambda,\Delta k)+\Delta S(\Lambda,\Delta k)}\,.
\ee  
To make our reasoning simpler, for the time being we suppose, that it is accurate enough to replace the summations on $\Delta k$ by integrals. This way we have 
\be
Z_L(\Lambda)=e^{-\beta Lf_{\rm min}(\Lambda)+\Delta S(\Lambda)}\,. 
\ee
Now
\be
-T\lim_{L\to\infty}{1\over L}\ln Z_L=f_{\rm min}(\Lambda)
\ee
from which the free energy density $f$ is obtained through a next limit
\be
f=\lim_{\Lambda\to\infty}f_{\rm min}(\Lambda)\,.
\ee
This leads to difficulties in filtering out the next to leading corrections, as 
\be
\lim_{L\to\infty}\ln\left(Z_L(\Lambda)e^{\beta Lf}\right)\to(+\mbox{or}-)\infty
\quad\quad\mbox{like}\quad\quad L(f-f_{\rm min}(\Lambda))\,,
\ee
i.e.~$f-f_{\rm min}(\Lambda)$ \emph{hides} the correction we want to get. (We have to emphasize, that the value of the saddle point corrections is not effected by this, nevertheless we have to see the leading order term more accurately than the correction we expect.)

To resolve this problem we propose the following. 
Taking larger and larger $L$ allows taking larger and larger $\Lambda$,
thus the two limits can be synchronized: a $\Lambda(L)$ can bee chosen so, that
\be
\Lambda(L)\mathrel{\mathop{\longrightarrow}_{L\to\infty}}\infty\,,
\ee
while the condition for applying Stirling's formula within the 
cutoffs is met, i.e. 
\be\label{acondition2}
L\Delta k\rho_0(\Lambda)\gg1\,.
\ee 
It seems plausible, that if it is possible  
\begin{itemize}
\item to choose $\Lambda(L)$ so, that $L(f-f_{\rm min}(\Lambda(L)))\to0$, 
\item and take also $\Delta k\to0$ so that the difference between the 
sums and integrals disappears,  
\end{itemize}
while (\ref{acondition2}) holds,
than the next to leading order correction to the free
energy is $\lim_{\Lambda\to\infty}\cS(\Lambda)$ indeed. 
 
In the following we argue, that for the repulsive $\delta$ Bose gas one can construct 
an appropriate cutoff procedure. First let us consider (\ref{acondition2}). Due to the BA 
equations 
\be
\rho_0(\Lambda)=\sigma{e^{-\beta\epsilon(\Lambda)}\over1+e^{-\beta\epsilon(\Lambda)}}
\left(1+O(K(\Lambda,0))\right)\sim\sigma e^{-\beta\epsilon(\Lambda)}\sim\sigma e^{-\beta e(\Lambda)}\,,
\ee
thus we require
\be\label{cond1}
L\Delta k\sigma e^{-\beta e(\Lambda)}\to\infty\,.
\ee
(Here we used also, that for large $\Lambda$ $\epsilon(\Lambda)=e(\Lambda)+O(K(\Lambda,0))$.)

The effect of introducing integrals instead of the sums can be estimated by an Euler-Maclaurin
type formula. We find the most significant part is
\be\label{errorsumvsint}
\sim L(\Delta k)^2{d\over d\Lambda}\sigma e^{-\beta\epsilon(\Lambda)}\,,
\ee
i.e.~we need
\be\label{cond2}
L(\Delta k)^2\sigma e^{-\beta e(\Lambda)}\Lambda\to0\,.
\ee
(We do not give details here, just note, that making the error due to (\ref{summationapproximation})
(Appendix \ref{sec:errorinfeef}) to disappear fast enough leads to the condition (\ref{cond2}) too.)

Finally we had to estimate $L(f-f_{\rm min}(\Lambda(L)))$ but for this we have to specify the 
procedure. One possibility is simply to omit all the modes outside the $\pm\Lambda$ interval
(corresponding to taking their energy equal to $\infty$), but in
this scheme $f_{\rm min}(\Lambda(L))$ does not converge in $\Lambda$ to $f$ fast enough.
A procedure providing a much faster convergence can be constructed realizing
that the particles of high energy behave as free ones. In this scheme the free energy of the system 
is built up of two parts: the contribution of the modes within the $\pm\Lambda$ interval is calculated using Stirling's formula (just as in the bulk of the paper), while the contribution of the modes outside the cutoffs is approximated by the contribution of free particles of energy $e(k)$
with density of states $\rho(k)+\rho_h(k)$ given by the BA equations.
In this scheme the minimization of the free energy leads to a dressed energy given by the
equation
\bea
\epsilon_{\Lambda}(k)&=&e(k)-{T}\int_{-\Lambda}^{\Lambda}
K(k^{\prime},k)\ln\left(1+
e^{-\beta\epsilon_{\Lambda}(k^{\prime})}\right)dk^{\prime}\\
&-&{T}\left(\int_{-\infty}^{-\Lambda}+\int_{\Lambda}^{\infty}\right)
K(k^{\prime},k)\ln\left(1+
e^{-\beta e(k^{\prime})}\right)dk^{\prime}\,,\nn
\eea
and the free energy density is 
\be
f_{\rm min}(\Lambda)=
-\int_{-\Lambda}^{\Lambda}
\ln\left(1+e^{-\beta\epsilon_{\Lambda}(k)}\right)\sigma dk
-\left(\int_{-\infty}^{-\Lambda}+\int_{\Lambda}^{\infty}\right)
\ln\left(1+e^{-\beta e(k)}\right)\sigma dk\,.
\ee
The leading part of $L(f-f_{\rm min}(\Lambda(L)))$ is of the order of
\be
\sim L\int_{\Lambda}^{\infty}\left(\ln\left(1+e^{-\beta e(k)}\right)-
\ln\left(1+e^{-\beta\epsilon(k)}\right)\right)\sigma dk\,,
\ee
what through some straightforward manipulations and approximations leads to the condition
\be\label{cond3}
Le^{-\beta e(\Lambda)}{\sigma\over\Lambda^3}\to0\,.
\ee
It is not hard to see, that it is possible to define an 
$L\to\infty$, $\Lambda\to\infty$ and $\Delta k\to0$ limit so, that all the conditions 
(\ref{cond1})(\ref{cond2}) and (\ref{cond3}) are satisfied.

We have to note, that the above reasoning concerning the existence of appropriate cutoff procedure works for the 
$\delta$ Bose gas only, for other systems it has to be reformulated, but for certain models it is also possible, that
there is no need for such a cutoff procedure.

\section{}\label{sec:surfacepotential}

In the present Appendix we discus a case of the $\delta$ Bose gas with open ends, 
in which surface bound states can be present. The system is described by the Bethe Ansatz
equations
\be\label{LiLiemodappendix1}
2Lk_j=2\pi J_j-\varphi_0(k_j)-\varphi_L(k_j)-
\sum_{l(\not=j)}^N \left\{2\atn\left({k_j-k_l\over c}\right)
+ 2\atn\left({k_j+k_l\over c}\right)\right\}
\,.
\ee 
Now we suppose, $\varphi_L=\pi$ corresponding to an infinitely high wall closing the chain at
$L$, but we take
\be
\varphi_0(k)=\pi-\atn{k\over\gamma}\,.
\ee
Also this corresponds to an infinitely high wall, but this wall is preceded by an infinitely deep,
but also infinitely narrow potential well. An appropriate tuning of the width and depth of the
well leads to the above reflection phase shift. (The effect of such a potential in case of a 
$\delta$ Fermi gas is discussed in \cite{Wo}, and a similar case of a Hubbard chain in \cite{BedFra})
This potential can always generate at least one surface bound state: it is not hard to see, that
for any distribution of the real $k$s (we denote them by Latin $k$s) (\ref{LiLiemodappendix1})
has also imaginary solution too corresponding to a surface bound state at the end at $x=0$. 
Denoting this by $i\kappa$ we find 
\be 
\kappa=\gamma-\delta\,,\quad\quad \delta=2\gamma e^{-2L\gamma}
\prod_i {(c-\gamma)^2+k_i^2\over (c+\gamma)^2+k_i^2}
\ee
In the presence of such an imaginary wavenumber the real ones satisfy the equations
\be\label{LiLiemodappendix2}
2Lk_j=2\pi J_j+\atn{k_j\over\gamma}-t(k_j,\kappa)
-\sum_{l(\not=j)}^N \left\{2\atn\left({k_j-k_l\over c}\right)
+ 2\atn\left({k_j+k_l\over c}\right)\right\}
\,,
\ee
with
\be
t(k,\kappa)=\left\{2\atn\left({k-i\kappa\over c}\right)
+ 2\atn\left({k+i\kappa\over c}\right)\right\}\,.
\ee
If the potential is strong enough, i.e.~$\gamma>c$ 
(\ref{LiLiemodappendix1}) has also solutions with more than one imaginary wavenumbers. These describe
more than one particles bound to the $x=0$ end. For two imaginary wavenumbers $i\kappa_{1,2}$ we 
find
\bea
&&\kappa_1=\gamma-\delta_1\,,\quad\quad \delta_1\simeq2\gamma 
\left({\gamma-c\over \gamma}\right)^2
e^{-2L(2\gamma-c)}
\prod_i {(\gamma-c)^2+k_i^2\over (\gamma+c)^2+k_i^2}\,
{(\gamma-2c)^2+k_i^2\over (\gamma)^2+k_i^2}\nn\\
&&\\
&&\kappa_2=\kappa_1-c-\delta_2\,,\quad\quad \delta_2\simeq2c 
\left({\gamma-c\over \gamma}\right)
e^{-2L(\gamma-c)}
\prod_i 
{(\gamma-2c)^2+k_i^2\over (\gamma)^2+k_i^2}\,.\nn
\eea
In general, a solution with $\nu$ imaginary wavenumbers $i\kappa_{\alpha},\ \alpha=1,\ldots\nu$
exists, if $\gamma-(\nu-1)c>0$. The $i\kappa_{\alpha}$s have the form
\be
\kappa_1=\gamma-\delta_1\,,\quad\quad\kappa_{\alpha}=\kappa_{\alpha-1}-\delta_{\alpha}
\ \ (\alpha=2,\ldots\nu)
\ee
with all $\delta_{\alpha}$ being exponentially small in $L$.
In these solutions the real $k$ set satisfies the equations
\be\label{LiLiemodappendix3}
2Lk_j=2\pi J_j+\atn{k_j\over\gamma}-t_{\nu}(k_j)
-\sum_{l(\not=j)}^N \left\{2\atn\left({k_j-k_l\over c}\right)
+ 2\atn\left({k_j+k_l\over c}\right)\right\}
\,,
\ee
with
\be
t_{\nu}(k)=\sum_{\alpha=1}^{\nu}t(k,\kappa_{\alpha})
\,.
\ee
The thermodynamic treatment follows the procedure described in the bulk of the paper 
with the difference, that $\varphi_0(k)+\varphi_L(k)$ should be replaced by 
$-2\atn(k/\gamma)+t_{\nu}(k)$, and the calculation should be repeated for all possible $\nu$.
Now the minimal free energy at a given $\nu$ (taking also into account the direct energy contributions of the imaginary wavenumbers) is 
\be
\OL F_{\nu}=F_{\rm min}+\Delta F+\Delta F_{\nu}
\ee
with $F_{\rm min}$ being the bulk value, $\Delta F$ given by (\ref{DeltaF}), and
\be
\Delta F_{\nu=0}= {T\over2\pi}\int_0^{\infty}{2\gamma\over \gamma^2+k^2}\ln\left(1+e^{-\beta\epsilon(k)}\right)dk\,,
\ \ \ \ \Delta F_{\nu>0}=\Delta F_{0}+\sum_{\alpha=1}^{\nu}\epsilon(i\kappa_{\alpha})\,,
\ee
with $\epsilon(i\kappa)$ being the formal extension of (\ref{epsilon}) to complex $k$.
As the saddle point contribution is independent of the state of the surface finally we arrive at
\be
\OL Z=\sum_{\nu=0}^ne^{-\beta\OL F_{\nu}+\cS}=
e^{-\beta(F_{\rm min}+\Delta F)+\cS}\sum_{\nu=0}^ne^{-\beta\Delta F_{\nu}}
\ee
with $n$ being the maximal possible value of $\nu$ (that is always larger than or equal to one).

\section{}\label{sec:ln}

In this appendix we try to estimate the effect of the 
\be\label{extraf}
-T\ln\left(2\cL+1\right)
\ee
term in the free energy of an isotropic Heisenberg chain in no magnetic field. As $\cL$
is the magnetization of the highest weight member of the multiplet (which is the one described
by the BA equations)
\be
\cL={N\over2}-N\sum_{(n)}l_{(n)}\sum\rho^{(n)}(k)\Delta k\,.
\ee
For the densities $\rho^{(n)}_0$ given by the minimization of the leading part of the free
energy functional this is zero, thus
\be
\cL=-N\sum_{(n)}l_{(n)}\sum r^{(n)}(k)\Delta k\,,
\ee
with 
\be
r^{(n)}(k)=\rho^{(n)}(k)-\rho_0^{(n)}(k)\,.
\ee
In analogy with (\ref{xik}) new variables are introduced
\be\label{xikn}
\xi^{(n)}(k)=\sqrt{N\Delta k
{1\over2}{\rho^{(n)}_{0}(k)+\rho^{(n)}_{h,0}(k)\over
\rho^{(n)}_{0}(k)\rho^{(n)}_{h,0}(k)}}r^{(n)}(k)
\ee
leading to 
\be
\cL=-\sum_{(n)}l_{(n)}\sum \sqrt{2N\Delta k {\rho^{(n)}_{0}(k)\rho^{(n)}_{h,0}(k)\over
\rho^{(n)}_{0}(k)+\rho^{(n)}_{h,0}(k)}}\xi^{(n)}(k)\,.
\ee
As the application of Stirling's formula is correct if
\be
N\Delta k \rho^{(n)}_{0}(k)\gg1\quad{\rm and}\quad
N\Delta k \rho^{(n)}_{h,0}(k)\gg1\,,
\ee
-- just as in the case of free Fermi or $\delta$ Bose gas -- one has to make sure 
through a cutoff procedure, that these inequalities hold. As the main contribution of the 
saddle point fluctuations come from the $|\xi^{(n)}(k)|\sim O(1)$ region we may conclude
that for the thermodynamically important states $\cL$ can be large. 

The contribution of the (\ref{extraf}) can be estimated as follows.
Denoting the leading part of the free energy of the system as a function of the magnetization by $\textbf{F}(S)$,
the corrected free energy
\be
\textbf{F}(\cL)-T\ln(2\cL+1) 
\ee
has to be minimized. This leads to
\be
\cL\simeq\sqrt{T\chi}\,,
\ee
where the susceptibility
\be
\chi=\left(\left.{\partial^2\textbf{F}(S)\over\partial S^2}\right|_{S=0}\right)^{-1}\propto N\,.
\ee
This way the minimal value is 
\be
\textbf{F}(0)-T\ln(2\sqrt{T\chi}+1) 
\ee
indicating that the next to leading term is $O(\ln N)$, what is much larger than the 
$O(1)$ contribution of the saddle point fluctuations.



\begin{thebibliography}{99}
\bibitem{muinYY}
In the treatment of Yang and Yang the chemical potential is not added to the energy at the beginning,
but it enters the procedure when minimizing the free energy as the Lagrange multiplier connected to the condition $N={\rm const}$. In our calculation we have to lift this condition, thus we have to use the chemical potential $\mu$ as a variable instead of the particle number $N$.
\end{thebibliography}
\end{document}